\documentclass{bmcart}

\usepackage[T1]{fontenc}
\usepackage[utf8]{inputenc}

\usepackage{csquotes}
\usepackage[english]{babel}

\usepackage{natbib}
\usepackage[acronym,hyperfirst=false,nohypertypes={acronym}]{glossaries}

\usepackage{amsmath}
\usepackage{amssymb}
\usepackage{amsfonts}
\usepackage{microtype}
\usepackage{listings}

\usepackage{tabu}
\usepackage{xcolor}
\usepackage{graphicx}
\usepackage{booktabs}

\usepackage{hyperref}
\usepackage{cleveref}

\usepackage{todonotes}

\newcommand\YAMLcolonstyle{\color{red}\mdseries}
\newcommand\YAMLkeystyle{\color{black}\bfseries}
\newcommand\YAMLvaluestyle{\color{blue}\mdseries}

\renewcommand{\arraystretch}{1.2}

\graphicspath{{./figures/}}

\crefname{figure}{figure}{figures}
\crefname{figure}{Figure}{Figures}

\newacronym[shortplural={CPS}]{CPS}{CPS}{cyber-physical system}
\newacronym{AI}{AI}{Artificial Intelligence}
\newacronym{AL}{AL}{Adversarial Learning}
\newacronym{ANN}{ANN}{Artificial Neural Network}
\newacronym{ARL}{ARL}{Adversarial Resilience Learning}
\newacronym{GAN}{GAN}{Generative Adversarial Network}
\newacronym{GPGPU}{GPGPU}{General-Purpose Graphics Processing Unit}
\newacronym{DL}{DL}{Deep Learning}
\newacronym{GRU}{GRU}{Gated Recurrent Unit}
\newacronym{ICT}{ICT}{Information and Communication Technology}
\newacronym{LSTM}{LSTM}{Long-Short Term Memory}
\newacronym{AEG}{AEG}{Attack Execution Graph}
\newacronym{ML}{ML}{Machine Learning}
\newacronym{PoC}{PoC}{Proof-of-Concept}
\newacronym{RL}{RL}{Reinforcement Learning}
\newacronym{DNN}{DNN}{Deep Neural Network}
\newacronym{RNN}{RNN}{Recurrent Neural Network}
\newacronym{CNN}{CNN}{Convolutional Neural Network}
\newacronym{SIMD}{SIMD}{Single Instruction, Multiple Data}
\newacronym{DoE}{DoE}{design of experiments}
\newacronym{DNC}{DNC}{Differentiable Neural Computer}
\newacronym{PV}{PV}{Photovoltaic}
\newacronym{MV}{MV}{medium voltage}
\newacronym{CHP}{CHP}{combined heat and power}
\newacronym{STL}{STL}{Signal Temporal Logic}
\newacronym{MTL}{MTL}{Metric Temporal Logic}
\newacronym{MITL}{MITL}{Metric Interval Temporal Logic}
\newacronym{AV}{AV}{Autonomous Vehicle}
\newacronym[shortplural={MAS}]{MAS}{MAS}{Multi Agent System}
\newacronym{RTU}{RTU}{Remote Terminal Unit}
\newacronym{TCP}{TCP}{Transmission Control Protocol}
\newacronym{LPEP}{LPEP}{Lightweight Power Exchange Protocol}
\newacronym{FMI}{FMI}{Functional Mock-Up Interface}
\newacronym{XML}{XML}{eXtensible Markup Language}
\newacronym{IoT}{IoT}{Internet of Things}
\newacronym{SIEM}{SIEM}{Security Information and Event Management}
\newacronym{APT}{APT}{Advanced Persistent Threats}
\newacronym{TAL}{TAL}{Thread Agents Library}
\newacronym{CBN}{CBN}{Causal Bayesian Network}
\newacronym{DER}{DER}{Distributed Energy Resource}
\newacronym{CIGRE}{CIGRE}{International Council on Large Electric Systems}

\newacronym{PYRATE}{PYRATE}{Polymorphic agents as cross-sectional software technology for the analysis of the operational safety of cyber-physical systems}

\newglossaryentry{resilience}{%
  name=resilience,
  description={The ability of a system to withstand unforseen, rare and
  potentially catastrophic events and to self-heal or improve in
  reaction to these events. Ideally resilience is increasing monotonously.}
}

\newcommand{\ARLBox}[2]{{\ensuremath\mathit{Box}\big\{ (#1), (#2) \big\}}}
\newcommand{\ARLDiscrete}[1]{\ensuremath{\mathit{Discrete}\{ #1 \}}}

\begin{document}

\begin{frontmatter}
  \begin{fmbox}
    \dochead{research}
    \title{%
      Analyzing Power Grid, ICT, and Market\\
      Without Domain Knowledge\\ 
      Using Distributed Artificial Intelligence}
    \author[addressref=offis, email={eric.veith@offis.de}]{%
      \inits{EV}\fnm{Eric MSP} \snm{Veith}}
    \author[addressref=offis, email={staphan.balduin@offis.de}]{%
      \inits{SB}\fnm{Stephan} \snm{Balduin}}
    \author[addressref=offis, email={nils.wenninghoff@offis.de}]{%
      \inits{NW}\fnm{Nils} \snm{Wenninghoff}}
    \author[addressref=offis, email={martin.troeschel@offis.de}]{%
      \inits{MT}\fnm{Martin} \snm{Tröschel}}
    \author[addressref=offis, email={lars.fischer@offis.de}]{%
      \inits{LF}\fnm{Lars} \snm{Fischer}}
    \author[addressref=ei, email={niesse@ei.uni-hannover.de}]{%
      \inits{AN}\fnm{Astrid} \snm{Nieße}}
    \author[addressref=ei, email={wolgast@ei.uni-hannover.de}]{%
      \inits{TWo}\fnm{Thomas} \snm{Wolgast}}
    \author[addressref=fri, email={richard.sethmann@hs-bremen.de}]{%
      \inits{RS}\fnm{Richard} \snm{Sethmann}}
    \author[addressref=fri, email={bastian.fraune@hs-bremen.de}]{%
      \inits{BF}\fnm{Bastian} \snm{Fraune}}
    \author[addressref=fri, email={torben.woltjen@hs-bremen.de}]{%
      \inits{TW}\fnm{Torben} \snm{Woltjen}}
    \address[id=offis]{%
      \orgname{Research Division Energy, 
        OFFIS -- Insitute for Information Technology}%
      \city{Oldenburg}%
      \cny{Germany}}
    \address[id=ei]{%
      \orgname{Department of Energy Informatics,
        Leibniz University Hannover}%
      \city{Hannover}%
      \cny{Germany}}
    \address[id=fri]{%
      \orgname{Department for Computer Networks and Information Security,
        Hochschule Bremen}%
      \city{Bremen}%
      \cny{Germany}}
  \end{fmbox}

  \begin{abstractbox}
    \begin{abstract}

            Modern \glspl{CPS}, such as our energy infrastructure, are becoming increasingly complex: An ever-higher share of \gls{AI}-based technologies use the \gls{ICT} facet of energy systems for operation optimization, cost efficiency, and to reach CO\textsubscript{2} goals worldwide. At the same time, markets with increased flexibility and ever shorter trade horizons enable the multi-stakeholder situation that is emerging in this setting. These systems still form critical infrastructures that need to perform with highest reliability. However, today's \gls{CPS} are becoming too complex to be analyzed in the traditional monolithic approach, where each domain, e.g., power grid and \gls{ICT} as well as the energy market, are considered as separate entities while ignoring dependencies and side-effects. To achieve an overall analysis, we introduce the concept for an application of distributed artificial intelligence as a self-adaptive analysis tool that is able to analyze the dependencies between domains in \gls{CPS} by attacking them. It eschews pre-configured domain knowledge, instead exploring the \gls{CPS} domains for emergent risk situations and exploitable loopholes in codices, with a focus on rational market actors that exploit the system while still following the market rules.

    \end{abstract}
  \end{abstractbox}
\end{frontmatter}

\glsresetall

\section{Introduction and Related Work}
\label{sec:introduction}

During the last two decades, the power grid has seen an enormous development in the adoption of \gls{ICT} on a large scale in order to facilitate the inclusion of advanced methodologies, including \gls{AI}-based approaches. This increases efficiency and flexibility, which ultimately allows a higher share of renewable energy sources in the grid. However, together with a proceeding decentralization and the inclusion of energy markets, the complexity of the overall system also increased, with different factors adding to it, e.g., prosumers directly selling their \gls{PV} power or new market-based concepts for ancillary service provisioning, which need to be implemented by 2021 as per EU regulations~\citep{EU19}. 

Decentralized generation and consumption has led to the emergence of decentralized grid operation and control paradigms, many of which feature independent software agents. These \glspl{MAS} exist for different tasks, e.g., to equalize real power generation and consumption, or to facilitate voltage control on local levels. A newer example of such a decentralized, specifically all-encompassing \gls{MAS} that is aimed at including a high share of volatile, renewable energy sources is the 
\emph{Universal Smart Grid Agent} system~\citep{veith2013lightweight,ruppert2014evolutionary,veith2017universal}.

Assuming that major \gls{IoT} trends will also influence the future power
grid, the comprehensive use of \gls{ICT} and \gls{AI} technologies
will, through their complexity, inevitably create an obstacle for a reliable
operation of the power grid \citep{hanseth2007risk,Sculley2014}. At least since the cyber attack on the power grid of the Ukraine in December 2015~\citep{case2016analysis,hamilton2016lights}, energy systems are recognized as valuable and vulnerable targets. Further attacks were seen in different stages with varying targets until 2017~\citep{reuters2017ukraine}. These attacks demonstrate how \gls{ICT} has a vital role in modern energy distribution networks. It needs to be reliable to ensure a stable power grid. However, due to the increasing ICT in modern power grids, the attack surface is getting bigger. Darknet marketplaces offer DDoS-as-a-Service and other attack-services for small money \citep{Crawley2016}, which demonstrates that security testing is getting more important in this special domain.

Research actively addresses the numerous challenges that arise from the increased complexity and, thus, new attack vectors the emerge not only in the energy domain, but all \glspl{CPS} in general. Among them are neural control falsification, e.g., through \gls{AL}~\citep{Pei2017,Gehr2018,Yaghoubi2019}, false data injection as attacks on state estimators~\citep{Teixeira2010,sandberg2010security,liu2011false,Gao2015a,hu2018state}, or utilizing compromised assets to actively damage the \gls{CPS}~\citep{Ju2018b}. 

In addition, a new type of attack has emerged in market-connected \gls{CPS} like energy systems: The attack as a side effect of economically rational behavior. Energy markets are highly regulated in all countries. The need for regulation directly follows from the energy systems' inherent dependability on a dedicated infrastructure, like power grids, gas and heat networks. With this kind of infrastructure, a natural monopoly is given. To ensure system stability while optimizing costs, market-based approaches are regulated to realize access to this infrastructure and system stability responsibility. The adaption of regulative frameworks is late by design: Once a loophole has been found, regulation is readjusted.  Even if no outright cyber-attack is staged, actors in the market might exploit loopholes while still conforming to the rules. There are a couple of known examples where this has been done and actually affected the power grid, e.g., in Germany with Inc-Dec Gaming against the zonal system with uniform pricing scheme~\citep{Hirth2018}, or in another case in Great Britain~\citep{Konstantinidis2015}. 

However, in a recent survey looking at \glspl{CPS} from the perspective of \gls{AI} research, we found that a large portion of research focuses on a safe inclusion of \gls{AI} technologies, such as Deep Learning or decentralized control through \gls{MAS} in critical infrastructures, but also emphasizes the gaps between almost fully analyzed, reliable \gls{CPS} and the complexity introduced by these techniques. Additionally, there is currently no systemic ana\-lysis approach that includes \gls{AI} technologies as the driver to explore and analyze unknown \glspl{CPS} for
safety
\citep{veith2019cpsanalysis}.
This survey can be seen as the main motivational background for this work: Traditional methods for analyzing the operational safety of a \gls{CPS} can only cover specific, partial aspects. Hence, we found extensive research into many different aspects of safe \gls{CPS} operation, but no approach for systemic testing of intra- and inter-domain relationships. From the point of the analysis, this causes a fragmentation of the whole system into islands. Aggregating subsystems also means that the effects of the interaction of components as well as the influence of market actors is not completely covered. This holds especially true for systemic vulnerabilities, in which isolated parameters are within nominal boundaries, but the overall system is being destabilized through emergent effects. On the basis of the challenges outlined above, we create an intelligent, cross-sectional software technology for analyzing complex \gls{CPS} in 
project PYRATE.
It analyzes complex \gls{CPS} with interdependent components autonomously, finding vulnerabilities leading to systemic failures. The core of the software technology to be developed is based on learning software agents that interact with a model---ideally a digital twin---of a \gls{CPS}, using the resulting system states as reinforcing feedback signals for full self-adaptivity to efficiently explore the search space of actions for destabilizing ones.

Our project works on two different levels: On a methodical level, we plan to develop a universal methodology to analyze weaknesses of arbitrary \gls{CPS} by finding successful attack strategies. On a practical level, we apply this methodology to an exemplary scenario containing a power system, an ancillary service market, and an \gls{ICT} system, to demonstrate possible applications and the effectiveness of the methodology.

The remainder of this paper is structured as follows: Due to didactic reasons, we first introduce the three environments of our demonstrator, explain major challenges in them, and describe the co-simulation setup. Afterwards, we follow with a description of our cross-domain learning \gls{MAS} that explores a \gls{CPS} in order to defeat it. The experimentation process that underpins any analysis of our technology is described next, followed by the post-run analysis that aims to isolate the minimal chain of actions that led to \gls{CPS} failure. Finally, we conclude with an outlook towards the realization. 

\section*{Environment Under Scrutiny: A Demonstrator}
\label{sec:environment}

In the research project, a power grid, an \gls{ICT} network, and a local ancillary service market are simultaneously subjected to analysis, since the goal is to analyze interdependent behavior. Since the analysis cannot be performed on real infrastructure for obvious reasons, simulation models of each of the different domains are being synchronized at run-time using a co-simulation approach.

\subsection*{Power System}

In this project's demonstrator, we focus on distribution grids to show the feasibility of the approach. Today's distribution grids lend themselves very well: They contain both, distributed large and aggregateable small loads, connect the major portion of \glspl{DER}, and are currently subject to large-scale \gls{ICT} inclusion as well as the development of local ancillary market concepts. Furthermore, they form the smallest meaningful, mostly self-contained environment that features a complex \gls{CPS} with a variety of outside influence factors such as volatile power generation from renewable energy sources.

For simulation and benchmark purposes on distribution grid level, a scenario-based benchmark environment was developed.  This benchmark environment incorporates a \gls{MV} grid  developed by the \gls{CIGRE}~\citep{rudion2006design, cigre2014benchmark}, time series data of one year in 15min resolution (e.g., for wind, solar radiation, or consumption) from a former research project
\emph{Smart Nord}~\citep{hofmann2015smart}
and different component models like \gls{PV} or \gls{CHP}.
 
\subsection*{Ancillary Service Market}

For current energy markets, regulation is mainly settled, though adaptions still can be seen quite often, e.g., for optimization reasons.  When implementing new energy markets, a whole new set of regulations is needed, though: There is a lot of activity in the implementation of regional energy markets and cell-based approaches, which are still in their infancy. Thus, we can expect many upcoming iterations on the regulation sets \cite{WeinhardtMCHHKO19}. This holds especially true for all kinds of ancillary service markets, e.g., reactive power or flexibility markets~\citep{LilliuVDR19,ChauXBE19}. 

In this context, even new problems arise: We found that, for grid-stabilizing ancillary service markets, regional actors and even private households could cooperatively induce problems into the grid to later get paid for eliminating these very problems. E.g., if we assume that the grid operator has to procure reactive power in a purely market-based way, private households could synchronize their load behavior in order to manipulate the local voltage level and to violate the voltage band. That forces the grid operator to announce a reactive power auction, in which generator agents would offer reactive power provision as ancillary service. Afterwards, the generator and household agents would divide profits and start a new attack. Regulation for such problems is not known at all, especially as this kind of malicious  behavior is difficult to detect and proof.

Our methodology will help to systematically investigate and understand such profit-driven attacks, which will in turn allow for better market designs. For this, a local auction-based reactive power market with simple rules will be implemented as incentive for profit-driven attacks. This will allow for better understanding of possible attack vectors for profit maximization. Later, systematic comparison with more sophisticated market designs and rules will enable insights which market rules increase resilience against which attack strategies. Finally, we hope to find market designs that minimize attacks and that maximize grid stability as well as detectability of such attacks.

\subsection*{ICT Simulation}

Distributed power units are equipped with \gls{ICT} to connect to wide-area networks. This enables operators to regulate and monitor distributed locations remotely, which is the foundation for implementing local ancillary service markets at the distribution grid level. The \gls{CIGRE} \gls{MV}~model only specifies a distribution grid topology without covering the \gls{ICT} domain, thus, we extend and overlay it with relevant \gls{ICT} components to model a realistic multi-domain distribution grid infrastructure. Consequently, each node of the energy grid is accompanied by the corresponding representation in the \gls{ICT} network that would, in reality, provide access to relevant sensors and actuators. Additionally, a communication network is built with routers and switches that connects these \glspl{CPS}, arranged in multiple subnets, hence modelling a realistic \gls{ICT} network.

Specific requirements arise from the multi-domain co-simulation setting: First, it needs to be efficient at simulating large networks. Second, the \gls{ICT} simulation is required to create an accurate model of the reality and, therefore, compute realistic results, which is especially important when examining networks in a security context. Thus, it is necessary that existing software can be integrated with minimal modifications. Lastly, the simulation tool needs to be easy-to-use, so that also experts of the other simulated domains---who might have limited knowledge about \gls{ICT} networks---can work with it after a short period of time. As there is no such simulator available that can meet all of these requirements, 
the \emph{rettij} network simulator was developed.
It is designed to simulate \gls{ICT} components like routers, switches, clients and servers, provided as Docker containers~\citep{dockerweb} in order to represent a realistic behaviour as opposed to synthetic, simulated models. The configuration files of the \gls{ICT} simulator integrate tightly with the rest of the software stack 
\citep{woltjen2020rettij}

\subsection*{Co-Simulation} 

The multi-domain simulation for analysis can hardly be performed by one software tool alone. The setup of the last three sections describes three different, but intertwined, domains; each one warrants its own specific simulation software to yield realistic results~
\citep{balduin2019towards}
In addition, specific models for power plants, wind parks, or independent market actors exist. These components are coordinated with the open-source co-simulation framework \emph{mosaik}~\citep{mosaikweb} and can therefore easily be integrated in other simulation setups relying on \emph{mosaik}. 

\begin{figure}
    \centering
    \includegraphics[width=0.9\textwidth]{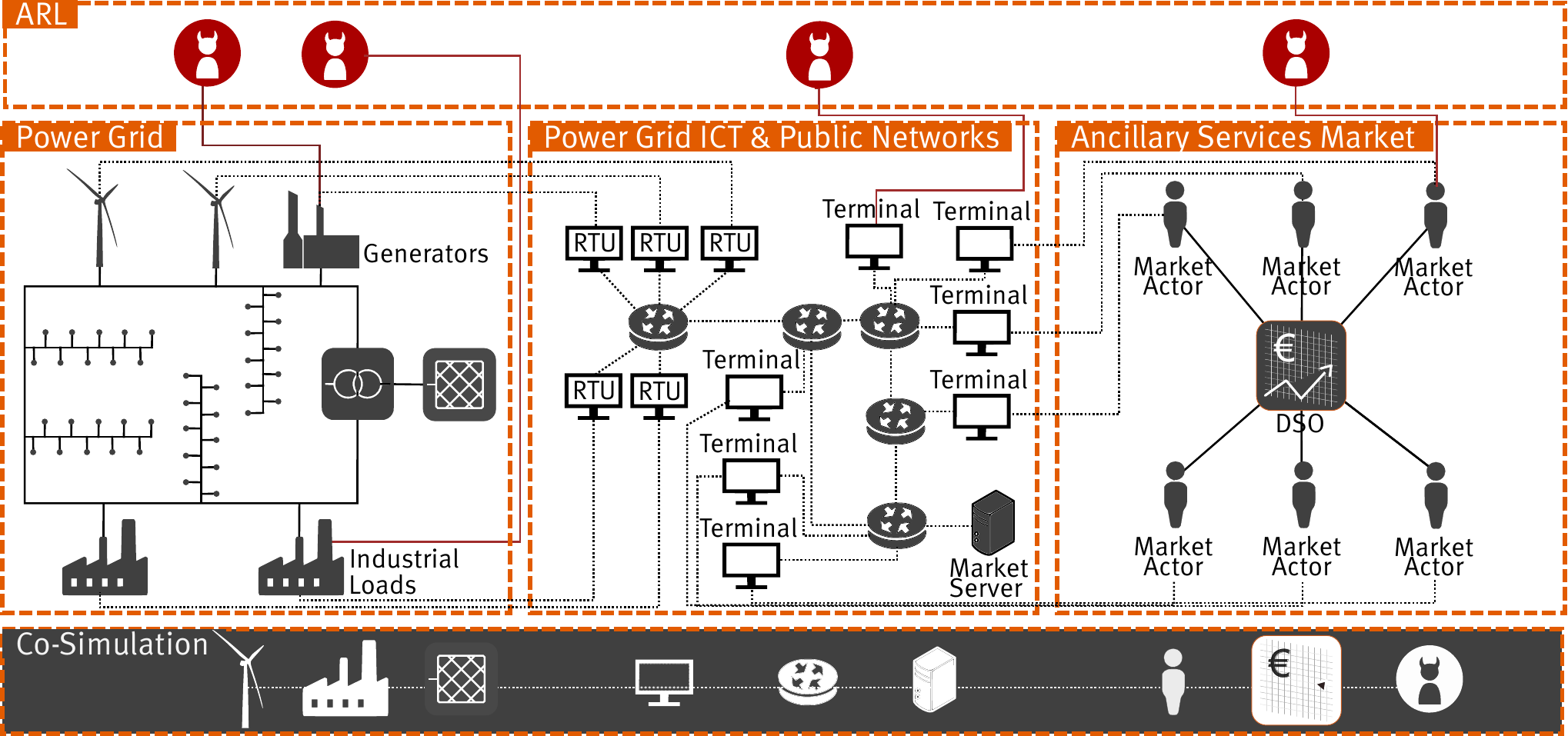}
    \caption{The Demonstrator's Co-Simulated Environments for Analysis} 
    \label{fig:pyrate-software-schema}
\end{figure}

\Cref{fig:pyrate-software-schema} shows the complete software stack. The bottom box, labelled \emph{co-simulation}, provides the technical view of the different simulators. Each simulator offers models as well as attributes on these models, which form a hierarchy: The address scheme \texttt{Simulator.Model.Attribute} allows for unambiguous identification of each individual attribute and to connect them. E.g., \texttt{ARL.Attacker-1.Actuator-1} can be connected to \texttt{PowerGrid.WindFarm-1.P-Feedin} to deliver setpoints from the adaptive attacker agent to a wind farm under its control; similarly, \texttt{PowerGrid.Sensor-1.Voltage}, connected to \texttt{ARL.Attacker-1.Sensor-1}, allows the agent to measure the effects of its actions in terms of voltage values. \emph{mosaik} synchronizes all simulators with each other and provides a common simulation clock time, the \emph{time step}; data is transmitted to a simulator when it is \emph{stepped}, data from its models' attributes is queried afterwards.

\section*{Distributed Analysis: Communication \& Control}
\label{sec:distributed-analysis}

To analyse this interconnected complex system, the core tool is the application of the \gls{ARL} methodology. \gls{ARL} defines in its pure form~
\citep{Fischer2019arl} 
two classes of
agents: \emph{attacker agents} and \emph{defender agents}. An instance of
every class operates on a model of a \gls{CPS}, i.e., both agents operate on
the same shared model. However, neither attacker nor defender know of each
other: They gather data from the \gls{CPS} through their sensors, which
retrieve the current state of the system---as far as it is observable to the
respective agent---, but do not explicitly track changes induced by another party.

This specific distinction makes sense for the power grid as well as for many
other \glspl{CPS}: Whether a voltage irregularity is induced by a larger
\gls{PV} feed-in at the end of the branch (e.g., coming from a farm) or forms
a part of an attack, is hardly distinguishable, but needs to be countered in any case.
Stringently, we assume that the defender needs to counter a variety of effects
for resilient operation, from fluctuation in renewable feed-in
to accidents to actual attacks without differentiating between them as a
rule-based system would do. Therefore, neither the overall system design nor
the experimenter differentiates between different causes and effects, leaving
the development of strategies as well as countering the adaption of the
attacker to the defender's capability to adapt (and vice versa). That both
agents learn to counter each other's strategies, thus developing them further
and further, is the core of the system-of-systems learning principle of
\gls{ARL}~\citep{veith2020adversarial}.
Consequently, we use the attacker not just to execute actual cyber-attacks, but to represent any potentially system-harming behavior. Thus, the attacker becomes a universal analysis tool.

Focusing on the attacker, we consider a group of attacking \gls{ARL} agents that form a self-organizing \gls{MAS} and a single defender agent that represents the grid operator. All \gls{ARL} agents use a modified \gls{RL} algorithm to explore a system that is initially unknown to them. In fact, \gls{ARL} agents possess no domain-specific knowledge; their sensors and actuators contain only a description of the space for valid values. For the experimenter, these space types provide an easy way to describe types and boundaries for values; they can also be used as predicates to check whether a concrete value is a valid member of the given space. E.g., for a given value \(x\), \(x\) is a member of the space \ARLDiscrete{x} iff:

\begin{equation}
  \ARLDiscrete{n}: x \in \mathbb{N}, 0 \le x \le n-1~.
\end{equation}

Similarly, we can denote a box in \(\mathbb{R}^n\) and check for a value \(x\) to be a member of it:

\begin{equation}
  \ARLBox{l_1, \dotsc, l_n}{h_1, \dotsc, h_n}:
  x \in \mathbb{R}, \bigwedge_{i=1}^{n} l_i \le x \le h_i~.
\end{equation}

Other space types are \emph{MultiDiscrete}, \emph{MultiBinary}, or \emph{Tuple}. Such a space description might represent the state of a tap-changer or the feed-in of a power plant in terms of a faction of its nominal output, but this logic is completely hidden from the agent. In fact, the domain logic is the responsibility of the experimenter. As the only way for \gls{RL} agents to learn is to receive feedback, the experimenter has to derive a proper reward function that covers the relevant aspects of the \gls{CPS}. The reward function bridges the otherwise separated concerns, i.e., the \gls{ARL} perspective that eschews domain knowledge and the \gls{CPS} domain.

Because of this feature, we describe the agents as being \emph{polymorphic}. Drawing from the analogy in software engineering, the agents' interfaces are fixed and soundly described, but do not carry any model. That means, the space types assigned to the agents' sensors and actuators form a declaration, but no definition. The agents derive this definition---i.e., their model---through exploration. Hence, they are polymorphic. This means that an abstract definition of a \gls{CPS}' interface in terms of the spaces outlined above is enough to have the \gls{ARL} agents explore the systems; this constitutes a fundamental difference from many modelling and analysis tools that require implicit or explicit modelling of the target domain.

As part of this new research direction, we assume that \glspl{MAS} are a valid approach to analyze highly decentralized systems as depicted above: They inherently allow for a representation of local knowledge and rule sets, even learned one, such as a limited view on local grid state and local control options~\citep{0022228}. It has already been shown that a combination with cyber-physical energy system simulation is feasible and beneficial to analyze the distributed behavior of the system, even for socio-technical system views~\citep{pra08}.  Thus, we use \glspl{MAS} to represent and explore the effect of cooperative malicious actors. In this case, cooperative means that the agents act cooperatively within their defined group of malicious or unplanned malicious, simply economically rational, agents. The attackers share a reward function, which can be as easy as the amount of money gained from the market, but also be complex and encompass aspects of all domains. In any case, the reward function remains transparent to each attacker and does not convey any domain knowledge to the agents, but is defined solely at the discretion of the experimenter.

In the presented concept, the overall \gls{MAS} encompasses all three domains.
Individual agents represent different actors in one of the domains. E.g., in a
scenario, in which the attacker \gls{MAS} controls three assets in the power grid, has one
entry point to the \gls{ICT} network, and appears with one bidder on
the market, the \gls{MAS} is comprised of five agents. An example for sensor and actuator mappings is presented in \cref{tab:sensor-actuator-mapping-example}.

{\def\arraystretch{1.5}
\begin{table}
    \newcommand{\ARLV}{Voltage\textsuperscript{1}}
    \newcommand{\ARLP}{Active Power\textsuperscript{2}}
    \newcommand{\ARLQ}{Reactive Power\textsuperscript{3}}
    \caption{Exemplary ARL Attacker MAS that can participate in a reactive power market}
    \label{tab:sensor-actuator-mapping-example}
    \begin{tabu}{XX[l,1.5]X[l,4]X[l,4]}
        \toprule
        \textbf{Agent} & \textbf{Asset} & \textbf{Sensors} & \textbf{Actuators}\\
        \midrule
        \(a_1\) & \gls{PV} Unit & \ARLV, Max. \ARLP & \ARLP, \ARLQ\\
        \(a_2\) & EV Charger & \ARLV, \ARLP & \ARLP, \ARLQ\\
        \(a_3\) & Load & \ARLV, \ARLP, \ARLQ & \ARLP\\
        \(a_4\) & Market & Reactive Power Commitment (relative)\textsuperscript{3} & Reactive Power Offer (relative)\textsuperscript{3}\\
        \(a_5\) & \gls{ICT} & Interface Utilization\textsuperscript{2} & Manipulate Sensor Value (Apply Noise)\textsuperscript{3}\\
        \bottomrule
    \end{tabu}\vspace{1ex}
    \raggedright\textsuperscript{1}$\ARLBox{0.85}{1.15}$, \textsuperscript{2}$\ARLBox{0.0}{1.0}$, \textsuperscript{3}$\ARLBox{-1.0}{1.0}$ 
\end{table}}

In order to develop an overall strategy, the attackers need to coordinate among themselves without a central command-\&-control instance. Snapshot algorithms \citep{cha85} will be used to enable the agents to interchange their local sensor data to gain knowledge of the global state. In this case, global state means the entirety of all sensors that the \gls{ARL} agents have access to. Learning agents that perform decision making based on shared knowledge can then learn optimal cooperative decision making based on that knowledge. With this research direction, we thrive for the development of a domain-encompassing coordination protocol to address this holistic approach to \gls{CPS} analysis.

While malicious cooperation cannot be deduced directly from regulatory or observability loopholes, beneficial cooperative behavior is analyzed as well: the defender aims to stabilize the system and prevent malicious attacks.

In our research approach, we therefore combine these agent types to act in shared environments. Thus, we hope to identify ruleset, \gls{ICT}, and market designs that minimize attack possibilities and stabilize the overall system. In future work, we will define and work out the resulting multi-layer attack coordination and defense framework.

\section*{Experiment Process}
\label{sec:experiment-process}

\begin{figure}
  \centering
  \includegraphics[width=0.9\textwidth]{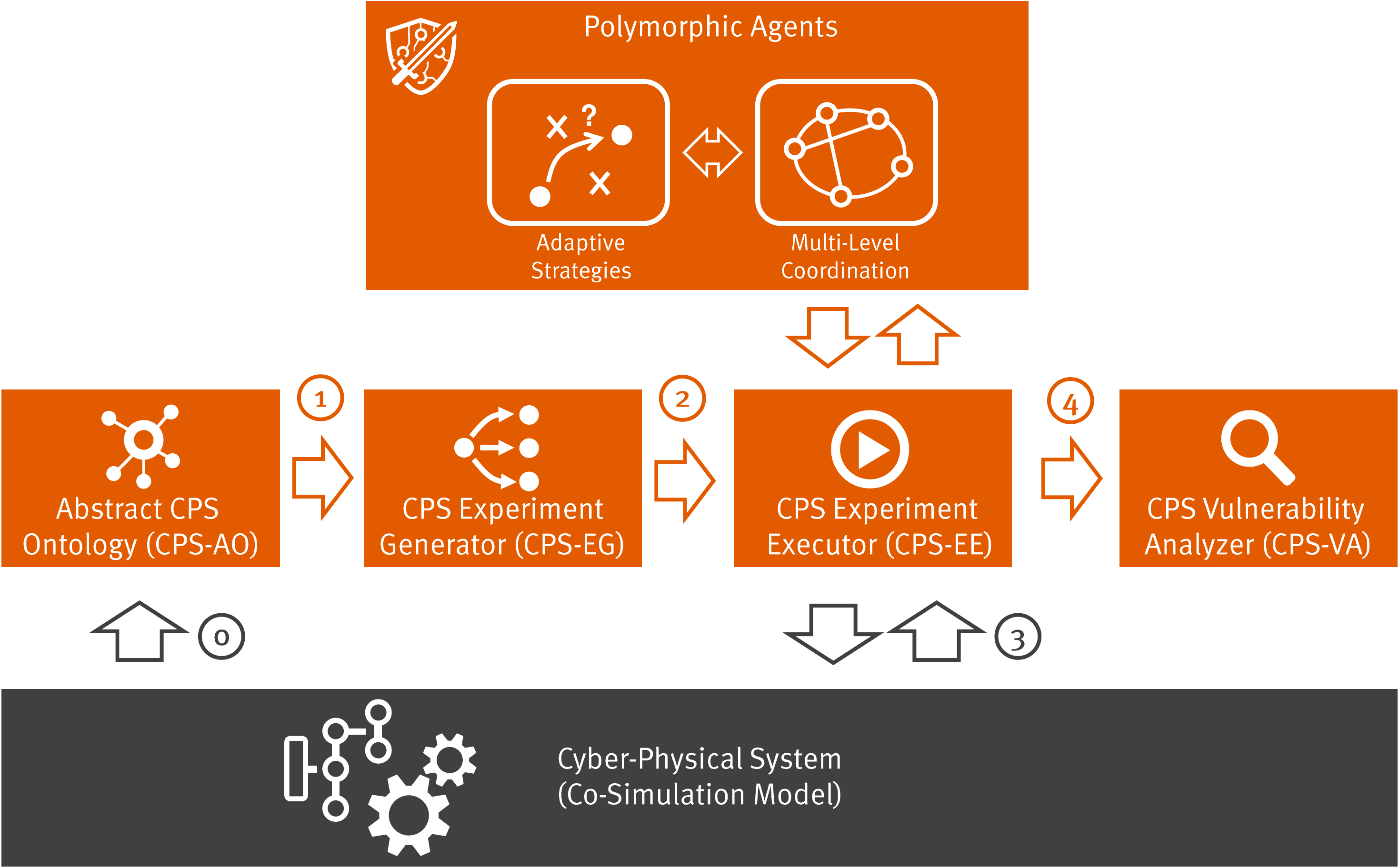}
  \caption{Experiment process of the presented research approach}
  \label{fig:expproc}
\end{figure}

As \cref{fig:expproc} illustrates, the overall experiment process incorporates four major steps: First, a domain independent description of the \gls{CPS} and its interfaces is required. The definition of such a description is called \emph{\gls{CPS} Abstract Ontology} (\gls{CPS}-AO) in the context of the presented research direction. The main purpose of the \gls{CPS}-AO is the definition of network topological variables and the mapping of the \gls{ARL} agents' sensors and actuators to entities in the environment. Additionally, the \gls{CPS}-AO defines which variables can be changed during the experiments and the valid value ranges. Furthermore, the \gls{CPS}-AO takes this topology information to build up experiments. For this purposes, \gls{CPS}-AO employs techniques from the domain of \gls{DoE}~\citep{kleijnen2015design} to select only configurations that provide the strongest significance. An example for a \gls{CPS}-AO configuration file can be seen in \cref{fig:cpsao-example}. 

\lstset{
  keywords={true,false,null,y,n},
  keywordstyle=\color{darkgray}\bfseries,
  basicstyle=\footnotesize\YAMLkeystyle,                                 
  sensitive=false,
  comment=[l]{\#},
  morecomment=[s]{/*}{*/},
  commentstyle=\color{purple}\ttfamily,
  stringstyle=\YAMLvaluestyle\ttfamily,
  moredelim=[l][\color{orange}]{\&},
  moredelim=[l][\color{magenta}]{*},
  moredelim=**[il][\YAMLcolonstyle{:}\YAMLvaluestyle]{:},   
  morestring=[b]',
  morestring=[b]",
  literate =    {---}{{\ProcessThreeDashes}}3
                {>}{{\textcolor{red}\textgreater}}1     
                {|}{{\textcolor{red}\textbar}}1 
                {\ -\ }{{\mdseries\ -\ }}3,
}
\begin{figure}
\begin{minipage}[t]{0.45\textwidth}
\begin{lstlisting}
!CPSAO
cps: !CPS # the system
  engine: # e.g. mosaik
  api: # how to instantiate
  sensors: # list of UIDs
  actuators: # list of UIDs
doe:
  runs: 
  factors: # DoE inputs 
  qualities: # DoE outputs
  strategy: # how to sample
  # e.g. Latin Hypercube
\end{lstlisting}
\end{minipage}
\begin{minipage}[t]{0.45\textwidth}
\begin{lstlisting}
agents: # two or more
- !Agent
# if more than one option
# is present in the agent's
# definition, they will be
# considered for DoE
  name: # convenience
  sensors: # list of UIDs
  actuators: # list of UIDs
  strategies: # how to win
  rewards: # 
- !Agent # same as above
\end{lstlisting}
\end{minipage}

\caption{A minimal example for a CPS-AO file. Additional parameters have been removed for the sake of brevity.}
\label{fig:cpsao-example}
\end{figure}

Furthermore, the \gls{CPS}-AO serves as an input for the so-called \emph{experiment generator} (\gls{CPS}-EG). While the \gls{CPS}-AO is a domain-independent and abstract description of the system, the \gls{CPS}-EG instantiates the experiment descriptions for the actual simulation, assigns values to factors, and builds execution scripts.

All so generated concrete experiments are executed by an \emph{experiment executor} (\gls{CPS}-EE) in the target environment. This provides the actual interface between the agent structure and the simulation environment.  In order to enable changes, the created intermediate results will  be saved so that smaller changes are possible without having to go through the entire process again.  During the execution of the experiments, the states of the simulation as well as the actions of the agents and the market results are stored and thus made available for a later weak point analysis, which will be described in the following section.

\section*{Post-Run Analysis Methodology}
\label{sec:post-run-analysis}

The experimenter defines a set of invariants that describe the environment's overall health. After the executor has finished the simulation run and health invariants were falsified, a \emph{post-mortem analysis} of the defeated system shall be conducted. This \gls{CPS} Vulnerability Analyzer (\gls{CPS}-VA) conducts targeted evaluation of the attacks across all domains, aiming to find the smallest chain of stringent actions that defeated this system, i.e. to identify the cross-domain attack-path or kill-chain. We assume that the \gls{ARL} \gls{MAS}, in its exploration, conducts a lot of negligible action before staging a successful attack. Hence, identification of the minimal kill-chain is a separate analysis task.

Another goal of the research project is the development of the \gls{CPS}-VA, which primarily aims to understand the produced data. It operates on data from all nodes, i.e. data from sensors and actuators. The transactions on the market as well as states from the \gls{ICT} and the power grid are collected. Then, the \gls{CPS}-VA is designed to apply different analysis techniques on this data to isolate the kill-chain. For understanding the path of the kill-chain, predictive models and techniques are often not the best choice \citep{Shmueli2010}. Most of the time, causality methods are more promising. Mueller, Memory, and Bartrem~\citep{Mueller2019} use causal discovery techniques to discover cyber kill-chains; therefore, data is presented as a \gls{CBN}. Finding the right methodology to explain the experiment's outcome is also part of the development of the \gls{CPS}-VA and will start as soon as first datasets are ready.

From the \gls{ICT} security point of view, the task of the \gls{CPS}-VA is somehow similar to what threat detection tools are designed for, but so far they only focus on \gls{ICT} related data. The behaviour of the \gls{ARL} agents can be treated as \gls{APT}. \Glspl{APT} can be described as sophisticated attack processes that are often strategically-motivated and profit-focused~\citep{Ahmad2019}. Standard industry solutions to detect \glspl{APT} are so called \gls{SIEM} systems. Such a system collects data from a wide variety of security applications to detect suspicious traffic and behaviour in \gls{ICT} systems. To make use of this information, \gls{SIEM} systems use correlation rules~\citep{Ambre2015} and rise alarm in case of a anomaly detection. The \gls{CPS}-VA provides the opportunity to evaluate the idea of \gls{SIEM} systems towards new applications.

First, a reasonable model from all domains is used in simulation to manually create simple correlation rules. This first step evaluates which information from the domains is necessary and how to create suitable correlation rules to generate basic knowledge for the next steps. Second, a much higher amount of relevant information for the \gls{SIEM} is expected. In correlating different
experiment runs from a variety of different scenarios---using, e.g., big data analytics~\citep{Cardenas2013}---, singular kill-chains can be derived and, thus, the respective rules be created. We expect that, starting with easy-to-observe critical states in the \glspl{CPS}, an isolation path beginning on the affected components in the \gls{CPS} can connect the critical states to market actors.

\section*{Outlook}
\label{sec:conclusion}

Many \gls{CPS} experience a broad addition of inputs, from self-driving capabilities over user inputs and \gls{IoT} technologies to a broad market adoption in the case of power systems. The emergence of complex \glspl{CPS} cannot be covered by traditional modelling and analysis techniques that can address only specific aspects of the overall system. In this paper, we proposed the concept for an application of distributed artificial intelligence as a self-adaptive analysis tool that is able to analyze the interdependencies between domains in \gls{CPS}, covering the whole system. It eschews pre-configured domain knowledge, instead exploring the \gls{CPS} domains for emergent risk situations and exploitable loopholes in codices, with a special focus on rational market actors that exploit the system while still following the rules of market.

In the future, we will demonstrate the feasibility of a cross-domain distributed analysis, documenting the experimentation system, the coordinating \gls{MAS}-based exploration tool as well as the analysis tool. With the latter, we aim to extract a reduced chain of actions leading to a cross-system exploitation, thereby isolating attack vectors and loopholes in codices. Furthermore, we expect the use of polymorphic agents to lead to new 
insights in the field of \gls{RL}. The \gls{ARL} agent interaction with the \gls{ICT}, which forms a central piece of the concept, will give new valuable insights of the \gls{ICT}'s critical role in modern \glspl{CPS}. This will enhance research towards new security tools for modern critical infrastructures.

\begin{backmatter}

    \section*{Acknowledgements}

    We would like to thank Sebastian Lehnhoff for his councel and valuable
    inputs.

    \section*{Funding}

    This work was funded by the German Federal Ministry of Education and Research through the project \emph{PYRATE} (01IS19021A).

    \section*{Availability of data and materials}

    Source code will be published after notification of acceptance.

    \section*{Author's contributions}

    All authors contributed equally to the paper.

    \section*{Competing interests}

    The authors declare that they have no competing interests.

    \bibliographystyle{bmc-mathphys}
    \bibliography{pyrate-note-paper.bib}


\begin{thebibliography}{43}
\ifx \bisbn   \undefined \def \bisbn  #1{ISBN #1}\fi
\ifx \binits  \undefined \def \binits#1{#1}\fi
\ifx \bauthor  \undefined \def \bauthor#1{#1}\fi
\ifx \batitle  \undefined \def \batitle#1{#1}\fi
\ifx \bjtitle  \undefined \def \bjtitle#1{#1}\fi
\ifx \bvolume  \undefined \def \bvolume#1{\textbf{#1}}\fi
\ifx \byear  \undefined \def \byear#1{#1}\fi
\ifx \bissue  \undefined \def \bissue#1{#1}\fi
\ifx \bfpage  \undefined \def \bfpage#1{#1}\fi
\ifx \blpage  \undefined \def \blpage #1{#1}\fi
\ifx \burl  \undefined \def \burl#1{\textsf{#1}}\fi
\ifx \doiurl  \undefined \def \doiurl#1{\textsf{#1}}\fi
\ifx \betal  \undefined \def \betal{\textit{et al.}}\fi
\ifx \binstitute  \undefined \def \binstitute#1{#1}\fi
\ifx \binstitutionaled  \undefined \def \binstitutionaled#1{#1}\fi
\ifx \bctitle  \undefined \def \bctitle#1{#1}\fi
\ifx \beditor  \undefined \def \beditor#1{#1}\fi
\ifx \bpublisher  \undefined \def \bpublisher#1{#1}\fi
\ifx \bbtitle  \undefined \def \bbtitle#1{#1}\fi
\ifx \bedition  \undefined \def \bedition#1{#1}\fi
\ifx \bseriesno  \undefined \def \bseriesno#1{#1}\fi
\ifx \blocation  \undefined \def \blocation#1{#1}\fi
\ifx \bsertitle  \undefined \def \bsertitle#1{#1}\fi
\ifx \bsnm \undefined \def \bsnm#1{#1}\fi
\ifx \bsuffix \undefined \def \bsuffix#1{#1}\fi
\ifx \bparticle \undefined \def \bparticle#1{#1}\fi
\ifx \barticle \undefined \def \barticle#1{#1}\fi
\ifx \bconfdate \undefined \def \bconfdate #1{#1}\fi
\ifx \botherref \undefined \def \botherref #1{#1}\fi
\ifx \url \undefined \def \url#1{\textsf{#1}}\fi
\ifx \bchapter \undefined \def \bchapter#1{#1}\fi
\ifx \bbook \undefined \def \bbook#1{#1}\fi
\ifx \bcomment \undefined \def \bcomment#1{#1}\fi
\ifx \oauthor \undefined \def \oauthor#1{#1}\fi
\ifx \citeauthoryear \undefined \def \citeauthoryear#1{#1}\fi
\ifx \endbibitem  \undefined \def \endbibitem {}\fi
\ifx \bconflocation  \undefined \def \bconflocation#1{#1}\fi
\ifx \arxivurl  \undefined \def \arxivurl#1{\textsf{#1}}\fi
\csname PreBibitemsHook\endcsname

\bibitem{EU19}
\begin{botherref}
\oauthor{\bsnm{{European Union}}}:
{DIRECTIVE (EU) 2019/944 OF THE EUROPEAN PARLIAMENT AND OF THE COUNCIL of 5
  June 2019 on common rules for the internal market for electricity and
  amending Directive 2012/27/EU}
(5 June 2019).
\url{https://eur-lex.europa.eu/legal-content/EN/TXT/PDF/?uri=CELEX:32019L0944}
\end{botherref}
\endbibitem

\bibitem{veith2013lightweight}
\begin{barticle}
\bauthor{\bsnm{Veith}, \binits{E.M.}},
\bauthor{\bsnm{Steinbach}, \binits{B.}},
\bauthor{\bsnm{Windeln}, \binits{J.}}:
\batitle{A lightweight distributed software agent for automatic demand—supply
  calculation in smart grids}.
\bjtitle{International Journal On Advances in Internet Technology}
\bvolume{7}(\bissue{1}),
\bfpage{97}--\blpage{113}
(\byear{2014}).
\bcomment{International Academy, Research, and Industry Association ({IARIA})}
\end{barticle}
\endbibitem

\bibitem{ruppert2014evolutionary}
\begin{bchapter}
\bauthor{\bsnm{Ruppert}, \binits{M.}},
\bauthor{\bsnm{Veith}, \binits{E.M.}},
\bauthor{\bsnm{Steinbach}, \binits{B.}}:
\bctitle{An evolutionary training algorithm for artificial neural networks with
  dynamic offspring spread and implicit gradient information}.
In: \bbtitle{The Sixth International Conference on Emerging Network
  Intelligence ({EMERGING 2014})}.
\bpublisher{IARIA XPS Press}, \blocation{???}
(\byear{2014}).
\bcomment{International Academy, Research, and Industry Association ({IARIA})}
\end{bchapter}
\endbibitem

\bibitem{veith2017universal}
\begin{bbook}
\bauthor{\bsnm{Veith}, \binits{E.M.}}:
\bbtitle{Universal Smart Grid Agent for Distributed Power Generation
  Management}.
\bpublisher{Logos Verlag Berlin GmbH},
\blocation{Berlin, Germany}
(\byear{2017})
\end{bbook}
\endbibitem

\bibitem{hanseth2007risk}
\begin{bbook}
\bauthor{\bsnm{Hanseth}, \binits{O.}},
\bauthor{\bsnm{Ciborra}, \binits{C.}}:
\bbtitle{Risk, Complexity and ICT}.
\bpublisher{Edward Elgar Publishing},
\blocation{Cheltenham, UK}
(\byear{2007})
\end{bbook}
\endbibitem

\bibitem{Sculley2014}
\begin{botherref}
\oauthor{\bsnm{Sculley}, \binits{D.}},
\oauthor{\bsnm{Holt}, \binits{G.}},
\oauthor{\bsnm{Golovin}, \binits{D.}},
\oauthor{\bsnm{Davydov}, \binits{E.}},
\oauthor{\bsnm{Phillips}, \binits{T.}},
\oauthor{\bsnm{Ebner}, \binits{D.}},
\oauthor{\bsnm{Chaudhary}, \binits{V.}},
\oauthor{\bsnm{Young}, \binits{M.}}:
{Machine Learning: The High Interest Credit Card of Technical Debt}.
SE4ML: Software Engineering for Machine Learning (NIPS 2014 Workshop),
1--9
(2014)
\end{botherref}
\endbibitem

\bibitem{case2016analysis}
\begin{botherref}
\oauthor{\bsnm{Case}, \binits{D.U.}}:
Analysis of the cyber attack on the ukrainian power grid.
Electricity Information Sharing and Analysis Center (E-ISAC)
(2016)
\end{botherref}
\endbibitem

\bibitem{hamilton2016lights}
\begin{barticle}
\bauthor{\bsnm{Styczynski}, \binits{J.}},
\bauthor{\bsnm{Beach-Westmoreland}, \binits{N.}}:
\batitle{When the lights went out: {Ukraine} cybersecurity threat briefing}.
\bjtitle{Booz Allen Hamilton}
\bvolume{12},
\bfpage{20}
(\byear{2016})
\end{barticle}
\endbibitem

\bibitem{reuters2017ukraine}
\begin{botherref}
\oauthor{\bsnm{Reuters}}:
Ukrainian banks, electricity firm hit by fresh cyber attack
(2017)
\end{botherref}
\endbibitem

\bibitem{Crawley2016}
\begin{botherref}
\oauthor{\bsnm{Crawley}, \binits{A.}}:
Hiring hackers.
Network Security
(9),
13--15
(2016)
\end{botherref}
\endbibitem

\bibitem{Pei2017}
\begin{botherref}
\oauthor{\bsnm{Pei}, \binits{K.}},
\oauthor{\bsnm{Cao}, \binits{Y.}},
\oauthor{\bsnm{Yang}, \binits{J.}},
\oauthor{\bsnm{Jana}, \binits{S.}}:
{DeepXplore}: Automated whitebox testing of deep learning systems
(2017).
\arxivurl{1705.06640}
\end{botherref}
\endbibitem

\bibitem{Gehr2018}
\begin{bchapter}
\bauthor{\bsnm{Gehr}, \binits{T.}},
\bauthor{\bsnm{Mirman}, \binits{M.}},
\bauthor{\bsnm{Drachsler-Cohen}, \binits{D.}},
\bauthor{\bsnm{Tsankov}, \binits{P.}},
\bauthor{\bsnm{Chaudhuri}, \binits{S.}},
\bauthor{\bsnm{Vechev}, \binits{M.}}:
\bctitle{{AI\textsuperscript{2}}: Safety and robustness certification of neural
  networks with abstract interpretation}.
In: \bbtitle{2018 IEEE Symposium on Security and Privacy (SP)},
pp. \bfpage{1}--\blpage{18}.
\bpublisher{IEEE}, \blocation{???}
(\byear{2018})
\end{bchapter}
\endbibitem

\bibitem{Yaghoubi2019}
\begin{botherref}
\oauthor{\bsnm{Yaghoubi}, \binits{S.}},
\oauthor{\bsnm{Fainekos}, \binits{G.}}:
Gray-box adversarial testing for control systems with machine learning
  components
\textbf{1}(1),
179--184
(2019).
\arxivurl{arXiv:1812.11958v1}
\end{botherref}
\endbibitem

\bibitem{Teixeira2010}
\begin{botherref}
\oauthor{\bsnm{Teixeira}, \binits{A.}},
\oauthor{\bsnm{Amin}, \binits{S.}},
\oauthor{\bsnm{Sandberg}, \binits{H.}},
\oauthor{\bsnm{Johansson}, \binits{K.H.}},
\oauthor{\bsnm{Sastry}, \binits{S.S.}}:
Cyber security analysis of state estimators in electric power systems.
Proceedings of the IEEE Conference on Decision and Control,
5991--5998
(2010)
\end{botherref}
\endbibitem

\bibitem{sandberg2010security}
\begin{bchapter}
\bauthor{\bsnm{Sandberg}, \binits{H.}},
\bauthor{\bsnm{Teixeira}, \binits{A.}},
\bauthor{\bsnm{Johansson}, \binits{K.H.}}:
\bctitle{On security indices for state estimators in power networks}.
In: \bbtitle{First Workshop on Secure Control Systems (SCS), Stockholm, 2010}
(\byear{2010})
\end{bchapter}
\endbibitem

\bibitem{liu2011false}
\begin{barticle}
\bauthor{\bsnm{Liu}, \binits{Y.}},
\bauthor{\bsnm{Ning}, \binits{P.}},
\bauthor{\bsnm{Reiter}, \binits{M.K.}}:
\batitle{False data injection attacks against state estimation in electric
  power grids}.
\bjtitle{ACM Transactions on Information and System Security (TISSEC)}
\bvolume{14}(\bissue{1}),
\bfpage{13}
(\byear{2011})
\end{barticle}
\endbibitem

\bibitem{Gao2015a}
\begin{bchapter}
\bauthor{\bsnm{Gao}, \binits{S.}},
\bauthor{\bsnm{Xie}, \binits{L.}},
\bauthor{\bsnm{Solar-Lezama}, \binits{A.}},
\bauthor{\bsnm{Serpanos}, \binits{D.}},
\bauthor{\bsnm{Shrobe}, \binits{H.}}:
\bctitle{Automated vulnerability analysis of ac state estimation under
  constrained false data injection in electric power systems}.
In: \bbtitle{Proceedings of the IEEE Conference on Decision and Control},
vol. \bseriesno{54},
pp. \bfpage{2613}--\blpage{2620}.
\bpublisher{IEEE}, \blocation{???}
(\byear{2015})
\end{bchapter}
\endbibitem

\bibitem{hu2018state}
\begin{barticle}
\bauthor{\bsnm{Hu}, \binits{L.}},
\bauthor{\bsnm{Wang}, \binits{Z.}},
\bauthor{\bsnm{Han}, \binits{Q.-L.}},
\bauthor{\bsnm{Liu}, \binits{X.}}:
\batitle{State estimation under false data injection attacks: Security analysis
  and system protection}.
\bjtitle{Automatica}
\bvolume{87},
\bfpage{176}--\blpage{183}
(\byear{2018})
\end{barticle}
\endbibitem

\bibitem{Ju2018b}
\begin{bchapter}
\bauthor{\bsnm{Ju}, \binits{P.}},
\bauthor{\bsnm{Lin}, \binits{X.}}:
\bctitle{Adversarial attacks to distributed voltage control in power
  distribution networks with {DERs}}.
In: \bbtitle{Proceedings of the Ninth International Conference on Future Energy
  Systems},
pp. \bfpage{291}--\blpage{302}
(\byear{2018}).
\bcomment{ACM}
\end{bchapter}
\endbibitem

\bibitem{Hirth2018}
\begin{botherref}
\oauthor{\bsnm{Hirth}, \binits{L.}},
\oauthor{\bsnm{Schlecht}, \binits{I.}}:
{Market-Based Redispatch in Zonal Electricity Markets}.
SSRN Electronic Journal
(055)
(2018)
\end{botherref}
\endbibitem

\bibitem{Konstantinidis2015}
\begin{botherref}
\oauthor{\bsnm{Konstantinidis}, \binits{C.}},
\oauthor{\bsnm{Strbac}, \binits{G.}}:
{Empirics of intraday and real-time markets in Europe: Great Britain}.
Technical report,
DIW – Deutsches Institut f{\"{u}}r Wirtschaftsforschung,
Berlin, Germany
(2015)
\end{botherref}
\endbibitem

\bibitem{veith2019cpsanalysis}
\begin{bchapter}
\bauthor{\bsnm{Veith}, \binits{E.M.}},
\bauthor{\bsnm{Fischer}, \binits{L.}},
\bauthor{\bsnm{Tr\"{o}schel}, \binits{M.}},
\bauthor{\bsnm{Nie\ss{}e}, \binits{A.}}:
\bctitle{Analyzing cyber-physical systems from the perspective of artificial
  intelligence}.
In: \bbtitle{Proceedings of the 2019 International Conference on Artificial
  Intelligence, Robotics and Control}.
\bsertitle{AIRC ’19},
pp. \bfpage{85}--\blpage{95}.
\bpublisher{Association for Computing Machinery},
\blocation{New York, NY, USA}
(\byear{2019}).
\burl{https://doi.org/10.1145/3388218.3388222}
\end{bchapter}
\endbibitem

\bibitem{rudion2006design}
\begin{bchapter}
\bauthor{\bsnm{Rudion}, \binits{K.}},
\bauthor{\bsnm{Orths}, \binits{A.}},
\bauthor{\bsnm{Styczynski}, \binits{Z.A.}},
\bauthor{\bsnm{Strunz}, \binits{K.}}:
\bctitle{Design of benchmark of medium voltage distribution network for
  investigation of dg integration}.
In: \bbtitle{2006 IEEE Power Engineering Society General Meeting},
p. \bfpage{6}
(\byear{2006}).
\bcomment{IEEE}
\end{bchapter}
\endbibitem

\bibitem{cigre2014benchmark}
\begin{bbook}
\bauthor{\bsnm{{CIGRE Task Force C6.04.02}}}:
\bbtitle{Benchmark Systems for Network Integration of Renewable and Distributed
  Energy Resources},
(\byear{2014})
\end{bbook}
\endbibitem

\bibitem{hofmann2015smart}
\begin{botherref}
\oauthor{\bsnm{Hofmann}, \binits{L.}},
\oauthor{\bsnm{Sonnenschein}, \binits{M.}}:
Smart nord -- final report.
Hartmann GmbH
(2015)
\end{botherref}
\endbibitem

\bibitem{WeinhardtMCHHKO19}
\begin{bchapter}
\bauthor{\bsnm{Weinhardt}, \binits{C.}},
\bauthor{\bsnm{Mengelkamp}, \binits{E.}},
\bauthor{\bsnm{Cramer}, \binits{W.}},
\bauthor{\bsnm{Hambridge}, \binits{S.}},
\bauthor{\bsnm{Hobert}, \binits{A.}},
\bauthor{\bsnm{Kremers}, \binits{E.}},
\bauthor{\bsnm{Otter}, \binits{W.}},
\bauthor{\bsnm{Pinson}, \binits{P.}},
\bauthor{\bsnm{Tiefenbeck}, \binits{V.}},
\bauthor{\bsnm{Zade}, \binits{M.}}:
\bctitle{How far along are local energy markets in the {DACH+} region?: {A}
  comparative market engineering approach}.
In: \bbtitle{Proceedings of the Tenth {ACM} International Conference on Future
  Energy Systems, e-Energy 2019, Phoenix, AZ, USA, June 25-28, 2019},
pp. \bfpage{544}--\blpage{549}.
\bpublisher{{ACM}}, \blocation{???}
(\byear{2019}).
\burl{https://doi.org/10.1145/3307772.3335318}
\end{bchapter}
\endbibitem

\bibitem{LilliuVDR19}
\begin{bchapter}
\bauthor{\bsnm{Lilliu}, \binits{F.}},
\bauthor{\bsnm{Vinyals}, \binits{M.}},
\bauthor{\bsnm{Denysiuk}, \binits{R.}},
\bauthor{\bsnm{Recupero}, \binits{D.R.}}:
\bctitle{A novel payment scheme for trading renewable energy in smart grid}.
In: \bbtitle{Proceedings of the Tenth {ACM} International Conference on Future
  Energy Systems, e-Energy 2019, Phoenix, AZ, USA, June 25-28, 2019},
pp. \bfpage{111}--\blpage{115}.
\bpublisher{{ACM}}, \blocation{???}
(\byear{2019}).
\burl{https://doi.org/10.1145/3307772.3328299}
\end{bchapter}
\endbibitem

\bibitem{ChauXBE19}
\begin{bchapter}
\bauthor{\bsnm{Chau}, \binits{S.C.}},
\bauthor{\bsnm{Xu}, \binits{J.}},
\bauthor{\bsnm{Bow}, \binits{W.}},
\bauthor{\bsnm{Elbassioni}, \binits{K.M.}}:
\bctitle{Peer-to-peer energy sharing: Effective cost-sharing mechanisms and
  social efficiency}.
In: \bbtitle{Proceedings of the Tenth {ACM} International Conference on Future
  Energy Systems, e-Energy 2019, Phoenix, AZ, USA, June 25-28, 2019},
pp. \bfpage{215}--\blpage{225}.
\bpublisher{{ACM}}, \blocation{???}
(\byear{2019}).
\burl{https://doi.org/10.1145/3307772.3328312}
\end{bchapter}
\endbibitem

\bibitem{dockerweb}
\begin{botherref}
\oauthor{\bsnm{{The Docker developers}}}:
Docker Website.
[Retrieved: 2020-05-26].
\url{https://www.docker.com/}
\end{botherref}
\endbibitem

\bibitem{woltjen2020rettij}
\begin{bchapter}
\bauthor{\bsnm{Woltjen}, \binits{T.}},
\bauthor{\bsnm{Gritzan}, \binits{G.}},
\bauthor{\bsnm{Kathmann}, \binits{P.}},
\bauthor{\bsnm{Sethmann}, \binits{R.}}:
\bctitle{{Simulationsumgebung für IKT-Netze zur Cyber-Abwehr}}.
In: \bbtitle{Tagungsband AALE 2020},
pp. \bfpage{233}--\blpage{239}.
\bpublisher{VDE Verlag}, \blocation{???}
(\byear{2020})
\end{bchapter}
\endbibitem

\bibitem{balduin2019towards}
\begin{barticle}
\bauthor{\bsnm{Balduin}, \binits{S.}},
\bauthor{\bsnm{Tr{\"o}schel}, \binits{M.}},
\bauthor{\bsnm{Lehnhoff}, \binits{S.}}:
\batitle{Towards domain-specific surrogate models for smart grid
  co-simulation}.
\bjtitle{Energy Informatics}
\bvolume{2}(\bissue{1}),
\bfpage{27}
(\byear{2019})
\end{barticle}
\endbibitem

\bibitem{mosaikweb}
\begin{botherref}
\oauthor{\bsnm{{The mosaik Developers}}}:
mosaik Website.
[Retrieved: 2020-05-26].
\url{https://mosaik.offis.de/}
\end{botherref}
\endbibitem

\bibitem{Fischer2019arl}
\begin{bchapter}
\bauthor{\bsnm{Fischer}, \binits{L.}},
\bauthor{\bsnm{Memmen}, \binits{J.-M.}},
\bauthor{\bsnm{Veith}, \binits{E.M.}},
\bauthor{\bsnm{Tr{\"o}schel}, \binits{M.}}:
\bctitle{Adversarial resilience learning---towards systemic vulnerability
  analysis for large and complex systems}.
In: \bbtitle{The Ninth International Conference on Smart Grids, Green
  Communications and IT Energy-aware Technologies (ENERGY 2019)},
vol. \bseriesno{9},
pp. \bfpage{24}--\blpage{32}
(\byear{2019})
\end{bchapter}
\endbibitem

\bibitem{veith2020adversarial}
\begin{botherref}
\oauthor{\bsnm{Veith}, \binits{E.M.}},
\oauthor{\bsnm{Wenninghoff}, \binits{N.}},
\oauthor{\bsnm{Frost}, \binits{E.}}:
The Adversarial Resilience Learning Architecture for AI-based Modelling,
  Exploration, and Operation of Complex Cyber-Physical Systems
(2020).
\arxivurl{2005.13601}
\end{botherref}
\endbibitem

\bibitem{0022228}
\begin{bbook}
\bauthor{\bsnm{Shoham}, \binits{Y.}},
\bauthor{\bsnm{Leyton{-}Brown}, \binits{K.}}:
\bbtitle{Multiagent Systems - Algorithmic, Game-Theoretic, and Logical
  Foundations}.
\bpublisher{Cambridge University Press},
\blocation{Cambridge, MA, USA}
(\byear{2009})
\end{bbook}
\endbibitem

\bibitem{pra08}
\begin{bchapter}
\bauthor{\bsnm{Praca}, \binits{I.}},
\bauthor{\bsnm{Morais}, \binits{H.}},
\bauthor{\bsnm{Ramos}, \binits{C.}},
\bauthor{\bsnm{Vale}, \binits{Z.}},
\bauthor{\bsnm{Khodr}, \binits{H.}}:
\bctitle{Multi-agent electricity market simulation with dynamic strategies {\&}
  virtual power producers}.
In: \bbtitle{2008 {IEEE} Power {\&} Energy Society General Meeting},
pp. \bfpage{1}--\blpage{8}.
\bpublisher{IEEE},
\blocation{Piscataway, NJ}
(\byear{2008})
\end{bchapter}
\endbibitem

\bibitem{cha85}
\begin{barticle}
\bauthor{\bsnm{Chandy}, \binits{K.M.}},
\bauthor{\bsnm{Lamport}, \binits{L.}}:
\batitle{{Distributed Snapshots: Determining Global States of Distributed
  Systems}}.
\bjtitle{{ACM Transactions on Computer Systems (TOCS)}}
\bvolume{3}(\bissue{1}),
\bfpage{63}--\blpage{75}
(\byear{1985})
\end{barticle}
\endbibitem

\bibitem{kleijnen2015design}
\begin{bchapter}
\bauthor{\bsnm{Kleijnen}, \binits{J.P.}}:
\bctitle{Design and analysis of simulation experiments}.
In: \bbtitle{International Workshop on Simulation},
pp. \bfpage{3}--\blpage{22}
(\byear{2015}).
\bcomment{Springer}
\end{bchapter}
\endbibitem

\bibitem{Shmueli2010}
\begin{barticle}
\bauthor{\bsnm{Shmueli}, \binits{G.}}:
\batitle{{To Explain or to Predict?}}
\bjtitle{Statistical Science}
\bvolume{25}(\bissue{3}),
\bfpage{289}--\blpage{310}
(\byear{2010}).
\arxivurl{1101.0891}
\end{barticle}
\endbibitem

\bibitem{Mueller2019}
\begin{botherref}
\oauthor{\bsnm{Mueller}, \binits{W.G.}},
\oauthor{\bsnm{Memory}, \binits{A.}},
\oauthor{\bsnm{Bartrem}, \binits{K.}}:
Causal discovery of cyber attack phases.
Proceedings - 18th IEEE International Conference on Machine Learning and
  Applications, ICMLA 2019,
1348--1352
(2019)
\end{botherref}
\endbibitem

\bibitem{Ahmad2019}
\begin{barticle}
\bauthor{\bsnm{Ahmad}, \binits{A.}},
\bauthor{\bsnm{Webb}, \binits{J.}},
\bauthor{\bsnm{Desouza}, \binits{K.C.}},
\bauthor{\bsnm{Boorman}, \binits{J.}}:
\batitle{{Strategically-motivated advanced persistent threat: Definition,
  process, tactics and a disinformation model of counterattack}}.
\bjtitle{Computers {\&} Security}
\bvolume{86},
\bfpage{402}--\blpage{418}
(\byear{2019})
\end{barticle}
\endbibitem

\bibitem{Ambre2015}
\begin{barticle}
\bauthor{\bsnm{Ambre}, \binits{A.}},
\bauthor{\bsnm{Shekokar}, \binits{N.}}:
\batitle{{Insider Threat Detection Using Log Analysis and Event Correlation}}.
\bjtitle{Procedia Computer Science}
\bvolume{45}(\bissue{C}),
\bfpage{436}--\blpage{445}
(\byear{2015})
\end{barticle}
\endbibitem

\bibitem{Cardenas2013}
\begin{barticle}
\bauthor{\bsnm{Cardenas}, \binits{A.a.}},
\bauthor{\bsnm{Manadhata}, \binits{P.K.}},
\bauthor{\bsnm{Rajan}, \binits{S.P.}}:
\batitle{Big data analytics for security}.
\bjtitle{IEEE Security {\&} Privacy}
\bvolume{11}(\bissue{6}),
\bfpage{74}--\blpage{76}
(\byear{2013})
\end{barticle}
\endbibitem

\end{thebibliography}

\newcommand{\BMCxmlcomment}[1]{}

\BMCxmlcomment{

<refgrp>

<bibl id="B1">
  <title><p>{DIRECTIVE (EU) 2019/944 OF THE EUROPEAN PARLIAMENT AND OF THE
  COUNCIL of 5 June 2019 on common rules for the internal market for
  electricity and amending Directive 2012/27/EU}</p></title>
  <aug>
    <au><cnm>{European Union}</cnm></au>
  </aug>
  <source>{Official Journal of the European Union}</source>
  <pubdate>5 June 2019</pubdate>
  <url>https://eur-lex.europa.eu/legal-content/EN/TXT/PDF/?uri=CELEX:32019L0944</url>
</bibl>

<bibl id="B2">
  <title><p>A lightweight distributed software agent for automatic
  demand—supply calculation in smart grids</p></title>
  <aug>
    <au><snm>Veith</snm><fnm>EM</fnm></au>
    <au><snm>Steinbach</snm><fnm>B</fnm></au>
    <au><snm>Windeln</snm><fnm>J</fnm></au>
  </aug>
  <source>International Journal On Advances in Internet Technology</source>
  <publisher>IARIA XPS Press</publisher>
  <pubdate>2014</pubdate>
  <volume>7</volume>
  <issue>1</issue>
  <fpage>97</fpage>
  <lpage>-113</lpage>
</bibl>

<bibl id="B3">
  <title><p>An Evolutionary Training Algorithm for Artificial Neural Networks
  with Dynamic Offspring Spread and Implicit Gradient Information</p></title>
  <aug>
    <au><snm>Ruppert</snm><fnm>M</fnm></au>
    <au><snm>Veith</snm><fnm>EM</fnm></au>
    <au><snm>Steinbach</snm><fnm>B</fnm></au>
  </aug>
  <source>The Sixth International Conference on Emerging Network Intelligence
  ({EMERGING 2014})</source>
  <publisher>IARIA XPS Press</publisher>
  <pubdate>2014</pubdate>
</bibl>

<bibl id="B4">
  <title><p>Universal Smart Grid Agent for Distributed Power Generation
  Management</p></title>
  <aug>
    <au><snm>Veith</snm><fnm>EM</fnm></au>
  </aug>
  <publisher>Berlin, Germany: Logos Verlag Berlin GmbH</publisher>
  <pubdate>2017</pubdate>
</bibl>

<bibl id="B5">
  <title><p>Risk, complexity and ICT</p></title>
  <aug>
    <au><snm>Hanseth</snm><fnm>O</fnm></au>
    <au><snm>Ciborra</snm><fnm>C</fnm></au>
  </aug>
  <publisher>Cheltenham, UK: Edward Elgar Publishing</publisher>
  <pubdate>2007</pubdate>
</bibl>

<bibl id="B6">
  <title><p>{Machine Learning: The High Interest Credit Card of Technical
  Debt}</p></title>
  <aug>
    <au><snm>Sculley</snm><fnm>D</fnm></au>
    <au><snm>Holt</snm><fnm>G</fnm></au>
    <au><snm>Golovin</snm><fnm>D</fnm></au>
    <au><snm>Davydov</snm><fnm>E</fnm></au>
    <au><snm>Phillips</snm><fnm>T</fnm></au>
    <au><snm>Ebner</snm><fnm>D</fnm></au>
    <au><snm>Chaudhary</snm><fnm>V</fnm></au>
    <au><snm>Young</snm><fnm>M</fnm></au>
  </aug>
  <source>SE4ML: Software Engineering for Machine Learning (NIPS 2014
  Workshop)</source>
  <pubdate>2014</pubdate>
  <fpage>1</fpage>
  <lpage>-9</lpage>
</bibl>

<bibl id="B7">
  <title><p>Analysis of the cyber attack on the Ukrainian power
  grid</p></title>
  <aug>
    <au><snm>Case</snm><fnm>DU</fnm></au>
  </aug>
  <source>Electricity Information Sharing and Analysis Center (E-ISAC)</source>
  <pubdate>2016</pubdate>
</bibl>

<bibl id="B8">
  <title><p>When The Lights Went Out: {Ukraine} Cybersecurity Threat
  Briefing</p></title>
  <aug>
    <au><snm>Styczynski</snm><fnm>J</fnm></au>
    <au><snm>Beach Westmoreland</snm><fnm>N</fnm></au>
  </aug>
  <source>Booz Allen Hamilton</source>
  <pubdate>2016</pubdate>
  <volume>12</volume>
  <fpage>20</fpage>
</bibl>

<bibl id="B9">
  <title><p>Ukrainian banks, electricity firm hit by fresh cyber
  attack</p></title>
  <aug>
    <au><cnm>Reuters</cnm></au>
  </aug>
  <pubdate>2017</pubdate>
</bibl>

<bibl id="B10">
  <title><p>Hiring hackers</p></title>
  <aug>
    <au><snm>Crawley</snm><fnm>A</fnm></au>
  </aug>
  <source>Network Security</source>
  <publisher>Elsevier Ltd</publisher>
  <pubdate>2016</pubdate>
  <issue>9</issue>
  <fpage>13</fpage>
  <lpage>-15</lpage>
  <url>https://linkinghub.elsevier.com/retrieve/pii/S1353485816300885</url>
</bibl>

<bibl id="B11">
  <title><p>{DeepXplore}: Automated Whitebox Testing of Deep Learning
  Systems</p></title>
  <aug>
    <au><snm>Pei</snm><fnm>K</fnm></au>
    <au><snm>Cao</snm><fnm>Y</fnm></au>
    <au><snm>Yang</snm><fnm>J</fnm></au>
    <au><snm>Jana</snm><fnm>S</fnm></au>
  </aug>
  <pubdate>2017</pubdate>
  <url>http://arxiv.org/abs/1705.06640{\%}0Ahttp://dx.doi.org/10.1145/3132747.3132785</url>
</bibl>

<bibl id="B12">
  <title><p>{AI\textsuperscript{2}}: Safety and Robustness Certification of
  Neural Networks with Abstract Interpretation</p></title>
  <aug>
    <au><snm>Gehr</snm><fnm>T</fnm></au>
    <au><snm>Mirman</snm><fnm>M</fnm></au>
    <au><snm>Drachsler Cohen</snm><fnm>D</fnm></au>
    <au><snm>Tsankov</snm><fnm>P</fnm></au>
    <au><snm>Chaudhuri</snm><fnm>S</fnm></au>
    <au><snm>Vechev</snm><fnm>M</fnm></au>
  </aug>
  <source>2018 IEEE Symposium on Security and Privacy (SP)</source>
  <publisher>IEEE</publisher>
  <pubdate>2018</pubdate>
  <fpage>1</fpage>
  <lpage>-18</lpage>
</bibl>

<bibl id="B13">
  <title><p>Gray-box adversarial testing for control systems with machine
  learning components</p></title>
  <aug>
    <au><snm>Yaghoubi</snm><fnm>S</fnm></au>
    <au><snm>Fainekos</snm><fnm>G</fnm></au>
  </aug>
  <pubdate>2019</pubdate>
  <volume>1</volume>
  <issue>1</issue>
  <fpage>179</fpage>
  <lpage>-184</lpage>
</bibl>

<bibl id="B14">
  <title><p>Cyber security analysis of state estimators in electric power
  systems</p></title>
  <aug>
    <au><snm>Teixeira</snm><fnm>A</fnm></au>
    <au><snm>Amin</snm><fnm>S</fnm></au>
    <au><snm>Sandberg</snm><fnm>H</fnm></au>
    <au><snm>Johansson</snm><fnm>KH</fnm></au>
    <au><snm>Sastry</snm><fnm>SS</fnm></au>
  </aug>
  <source>Proceedings of the IEEE Conference on Decision and Control</source>
  <pubdate>2010</pubdate>
  <fpage>5991</fpage>
  <lpage>-5998</lpage>
</bibl>

<bibl id="B15">
  <title><p>On security indices for state estimators in power
  networks</p></title>
  <aug>
    <au><snm>Sandberg</snm><fnm>H</fnm></au>
    <au><snm>Teixeira</snm><fnm>A</fnm></au>
    <au><snm>Johansson</snm><fnm>KH</fnm></au>
  </aug>
  <source>First Workshop on Secure Control Systems (SCS), Stockholm,
  2010</source>
  <pubdate>2010</pubdate>
</bibl>

<bibl id="B16">
  <title><p>False data injection attacks against state estimation in electric
  power grids</p></title>
  <aug>
    <au><snm>Liu</snm><fnm>Y</fnm></au>
    <au><snm>Ning</snm><fnm>P</fnm></au>
    <au><snm>Reiter</snm><fnm>MK</fnm></au>
  </aug>
  <source>ACM Transactions on Information and System Security (TISSEC)</source>
  <publisher>ACM</publisher>
  <pubdate>2011</pubdate>
  <volume>14</volume>
  <issue>1</issue>
  <fpage>13</fpage>
</bibl>

<bibl id="B17">
  <title><p>Automated vulnerability analysis of AC state estimation under
  constrained false data injection in electric power systems</p></title>
  <aug>
    <au><snm>Gao</snm><fnm>S</fnm></au>
    <au><snm>Xie</snm><fnm>L</fnm></au>
    <au><snm>Solar Lezama</snm><fnm>A</fnm></au>
    <au><snm>Serpanos</snm><fnm>D</fnm></au>
    <au><snm>Shrobe</snm><fnm>H</fnm></au>
  </aug>
  <source>Proceedings of the IEEE Conference on Decision and Control</source>
  <publisher>IEEE</publisher>
  <pubdate>2015</pubdate>
  <volume>54</volume>
  <fpage>2613</fpage>
  <lpage>-2620</lpage>
</bibl>

<bibl id="B18">
  <title><p>State estimation under false data injection attacks: Security
  analysis and system protection</p></title>
  <aug>
    <au><snm>Hu</snm><fnm>L</fnm></au>
    <au><snm>Wang</snm><fnm>Z</fnm></au>
    <au><snm>Han</snm><fnm>QL</fnm></au>
    <au><snm>Liu</snm><fnm>X</fnm></au>
  </aug>
  <source>Automatica</source>
  <publisher>Elsevier</publisher>
  <pubdate>2018</pubdate>
  <volume>87</volume>
  <fpage>176</fpage>
  <lpage>-183</lpage>
</bibl>

<bibl id="B19">
  <title><p>Adversarial Attacks to Distributed Voltage Control in Power
  Distribution Networks with {DERs}</p></title>
  <aug>
    <au><snm>Ju</snm><fnm>P</fnm></au>
    <au><snm>Lin</snm><fnm>X</fnm></au>
  </aug>
  <source>Proceedings of the Ninth International Conference on Future Energy
  Systems</source>
  <pubdate>2018</pubdate>
  <fpage>291</fpage>
  <lpage>-302</lpage>
</bibl>

<bibl id="B20">
  <title><p>{Market-Based Redispatch in Zonal Electricity Markets}</p></title>
  <aug>
    <au><snm>Hirth</snm><fnm>L</fnm></au>
    <au><snm>Schlecht</snm><fnm>I</fnm></au>
  </aug>
  <source>SSRN Electronic Journal</source>
  <pubdate>2018</pubdate>
  <issue>055</issue>
</bibl>

<bibl id="B21">
  <title><p>{Empirics of intraday and real-time markets in Europe: Great
  Britain}</p></title>
  <aug>
    <au><snm>Konstantinidis</snm><fnm>C</fnm></au>
    <au><snm>Strbac</snm><fnm>G</fnm></au>
  </aug>
  <publisher>Berlin, Germany</publisher>
  <pubdate>2015</pubdate>
  <fpage>21</fpage>
</bibl>

<bibl id="B22">
  <title><p>Analyzing Cyber-Physical Systems from the Perspective of Artificial
  Intelligence</p></title>
  <aug>
    <au><snm>Veith</snm><fnm>EM</fnm></au>
    <au><snm>Fischer</snm><fnm>L</fnm></au>
    <au><snm>Tr\"{o}schel</snm><fnm>M</fnm></au>
    <au><snm>Nie\ss{}e</snm><fnm>A</fnm></au>
  </aug>
  <source>Proceedings of the 2019 International Conference on Artificial
  Intelligence, Robotics and Control</source>
  <publisher>New York, NY, USA: Association for Computing Machinery</publisher>
  <series><title><p>AIRC ’19</p></title></series>
  <pubdate>2019</pubdate>
  <fpage>85–95</fpage>
  <url>https://doi.org/10.1145/3388218.3388222</url>
</bibl>

<bibl id="B23">
  <title><p>Design of benchmark of medium voltage distribution network for
  investigation of DG integration</p></title>
  <aug>
    <au><snm>Rudion</snm><fnm>K</fnm></au>
    <au><snm>Orths</snm><fnm>A</fnm></au>
    <au><snm>Styczynski</snm><fnm>ZA</fnm></au>
    <au><snm>Strunz</snm><fnm>K</fnm></au>
  </aug>
  <source>2006 IEEE Power Engineering Society General Meeting</source>
  <pubdate>2006</pubdate>
  <fpage>6</fpage>
  <lpage>-pp</lpage>
</bibl>

<bibl id="B24">
  <title><p>Benchmark Systems for Network Integration of Renewable and
  Distributed Energy Resources</p></title>
  <aug>
    <au><cnm>{CIGRE Task Force C6.04.02}</cnm></au>
  </aug>
  <pubdate>2014</pubdate>
</bibl>

<bibl id="B25">
  <title><p>Smart Nord -- Final Report</p></title>
  <aug>
    <au><snm>Hofmann</snm><fnm>L</fnm></au>
    <au><snm>Sonnenschein</snm><fnm>M</fnm></au>
  </aug>
  <source>Hartmann GmbH</source>
  <publisher>Hannover, Germany</publisher>
  <pubdate>2015</pubdate>
</bibl>

<bibl id="B26">
  <title><p>How far along are Local Energy Markets in the {DACH+} Region?: {A}
  Comparative Market Engineering Approach</p></title>
  <aug>
    <au><snm>Weinhardt</snm><fnm>C</fnm></au>
    <au><snm>Mengelkamp</snm><fnm>E</fnm></au>
    <au><snm>Cramer</snm><fnm>W</fnm></au>
    <au><snm>Hambridge</snm><fnm>S</fnm></au>
    <au><snm>Hobert</snm><fnm>A</fnm></au>
    <au><snm>Kremers</snm><fnm>E</fnm></au>
    <au><snm>Otter</snm><fnm>W</fnm></au>
    <au><snm>Pinson</snm><fnm>P</fnm></au>
    <au><snm>Tiefenbeck</snm><fnm>V</fnm></au>
    <au><snm>Zade</snm><fnm>M</fnm></au>
  </aug>
  <source>Proceedings of the Tenth {ACM} International Conference on Future
  Energy Systems, e-Energy 2019, Phoenix, AZ, USA, June 25-28, 2019</source>
  <publisher>{ACM}</publisher>
  <pubdate>2019</pubdate>
  <fpage>544</fpage>
  <lpage>-549</lpage>
  <url>https://doi.org/10.1145/3307772.3335318</url>
</bibl>

<bibl id="B27">
  <title><p>A novel payment scheme for trading renewable energy in smart
  grid</p></title>
  <aug>
    <au><snm>Lilliu</snm><fnm>F</fnm></au>
    <au><snm>Vinyals</snm><fnm>M</fnm></au>
    <au><snm>Denysiuk</snm><fnm>R</fnm></au>
    <au><snm>Recupero</snm><fnm>DR</fnm></au>
  </aug>
  <source>Proceedings of the Tenth {ACM} International Conference on Future
  Energy Systems, e-Energy 2019, Phoenix, AZ, USA, June 25-28, 2019</source>
  <publisher>{ACM}</publisher>
  <pubdate>2019</pubdate>
  <fpage>111</fpage>
  <lpage>-115</lpage>
  <url>https://doi.org/10.1145/3307772.3328299</url>
</bibl>

<bibl id="B28">
  <title><p>Peer-to-Peer Energy Sharing: Effective Cost-Sharing Mechanisms and
  Social Efficiency</p></title>
  <aug>
    <au><snm>Chau</snm><fnm>SC</fnm></au>
    <au><snm>Xu</snm><fnm>J</fnm></au>
    <au><snm>Bow</snm><fnm>W</fnm></au>
    <au><snm>Elbassioni</snm><fnm>KM</fnm></au>
  </aug>
  <source>Proceedings of the Tenth {ACM} International Conference on Future
  Energy Systems, e-Energy 2019, Phoenix, AZ, USA, June 25-28, 2019</source>
  <publisher>{ACM}</publisher>
  <pubdate>2019</pubdate>
  <fpage>215</fpage>
  <lpage>-225</lpage>
  <url>https://doi.org/10.1145/3307772.3328312</url>
</bibl>

<bibl id="B29">
  <title><p>Docker Website</p></title>
  <aug>
    <au><cnm>{The Docker developers}</cnm></au>
  </aug>
  <url>https://www.docker.com/</url>
  <note>[Retrieved: 2020-05-26]</note>
</bibl>

<bibl id="B30">
  <title><p>{Simulationsumgebung für IKT-Netze zur Cyber-Abwehr}</p></title>
  <aug>
    <au><snm>Woltjen</snm><fnm>T</fnm></au>
    <au><snm>Gritzan</snm><fnm>G</fnm></au>
    <au><snm>Kathmann</snm><fnm>P</fnm></au>
    <au><snm>Sethmann</snm><fnm>R</fnm></au>
  </aug>
  <source>Tagungsband AALE 2020</source>
  <publisher>VDE Verlag</publisher>
  <pubdate>2020</pubdate>
  <fpage>233</fpage>
  <lpage>239</lpage>
</bibl>

<bibl id="B31">
  <title><p>Towards domain-specific surrogate models for smart grid
  co-simulation</p></title>
  <aug>
    <au><snm>Balduin</snm><fnm>S</fnm></au>
    <au><snm>Tr{\"o}schel</snm><fnm>M</fnm></au>
    <au><snm>Lehnhoff</snm><fnm>S</fnm></au>
  </aug>
  <source>Energy Informatics</source>
  <publisher>Springer</publisher>
  <pubdate>2019</pubdate>
  <volume>2</volume>
  <issue>1</issue>
  <fpage>27</fpage>
</bibl>

<bibl id="B32">
  <title><p>mosaik Website</p></title>
  <aug>
    <au><cnm>{The mosaik Developers}</cnm></au>
  </aug>
  <url>https://mosaik.offis.de/</url>
  <note>[Retrieved: 2020-05-26]</note>
</bibl>

<bibl id="B33">
  <title><p>Adversarial Resilience Learning---Towards Systemic Vulnerability
  Analysis for Large and Complex Systems</p></title>
  <aug>
    <au><snm>Fischer</snm><fnm>L</fnm></au>
    <au><snm>Memmen</snm><fnm>JM</fnm></au>
    <au><snm>Veith</snm><fnm>EM</fnm></au>
    <au><snm>Tr{\"o}schel</snm><fnm>M</fnm></au>
  </aug>
  <source>The Ninth International Conference on Smart Grids, Green
  Communications and IT Energy-aware Technologies (ENERGY 2019)</source>
  <pubdate>2019</pubdate>
  <volume>9</volume>
  <issue>1</issue>
  <fpage>24</fpage>
  <lpage>-32</lpage>
</bibl>

<bibl id="B34">
  <title><p>The Adversarial Resilience Learning Architecture for AI-based
  Modelling, Exploration, and Operation of Complex Cyber-Physical
  Systems</p></title>
  <aug>
    <au><snm>Veith</snm><fnm>EM</fnm></au>
    <au><snm>Wenninghoff</snm><fnm>N</fnm></au>
    <au><snm>Frost</snm><fnm>E</fnm></au>
  </aug>
  <pubdate>2020</pubdate>
</bibl>

<bibl id="B35">
  <title><p>Multiagent Systems - Algorithmic, Game-Theoretic, and Logical
  Foundations</p></title>
  <aug>
    <au><snm>Shoham</snm><fnm>Y</fnm></au>
    <au><snm>Leyton{-}Brown</snm><fnm>K</fnm></au>
  </aug>
  <publisher>Cambridge, MA, USA: Cambridge University Press</publisher>
  <pubdate>2009</pubdate>
</bibl>

<bibl id="B36">
  <title><p>Multi-agent electricity market simulation with dynamic strategies
  {&amp} virtual power producers</p></title>
  <aug>
    <au><snm>Praca</snm><fnm>I</fnm></au>
    <au><snm>Morais</snm><fnm>H</fnm></au>
    <au><snm>Ramos</snm><fnm>C</fnm></au>
    <au><snm>Vale</snm><fnm>Z</fnm></au>
    <au><snm>Khodr</snm><fnm>H</fnm></au>
  </aug>
  <source>2008 {IEEE} Power {\&} Energy Society general meeting</source>
  <publisher>Piscataway, NJ: IEEE</publisher>
  <pubdate>2008</pubdate>
  <fpage>1</fpage>
  <lpage>-8</lpage>
</bibl>

<bibl id="B37">
  <title><p>{Distributed Snapshots: Determining Global States of Distributed
  Systems}</p></title>
  <aug>
    <au><snm>Chandy</snm><fnm>KM</fnm></au>
    <au><snm>Lamport</snm><fnm>L</fnm></au>
  </aug>
  <source>{ACM Transactions on Computer Systems (TOCS)}</source>
  <pubdate>1985</pubdate>
  <volume>3</volume>
  <issue>1</issue>
  <fpage>63</fpage>
  <lpage>-75</lpage>
</bibl>

<bibl id="B38">
  <title><p>Design and analysis of simulation experiments</p></title>
  <aug>
    <au><snm>Kleijnen</snm><fnm>JP</fnm></au>
  </aug>
  <source>International Workshop on Simulation</source>
  <pubdate>2015</pubdate>
  <fpage>3</fpage>
  <lpage>-22</lpage>
</bibl>

<bibl id="B39">
  <title><p>{To Explain or to Predict?}</p></title>
  <aug>
    <au><snm>Shmueli</snm><fnm>G</fnm></au>
  </aug>
  <source>Statistical Science</source>
  <pubdate>2010</pubdate>
  <volume>25</volume>
  <issue>3</issue>
  <fpage>289</fpage>
  <lpage>-310</lpage>
  <url>http://projecteuclid.org/euclid.ss/1294167961</url>
</bibl>

<bibl id="B40">
  <title><p>Causal discovery of cyber attack phases</p></title>
  <aug>
    <au><snm>Mueller</snm><fnm>WG</fnm></au>
    <au><snm>Memory</snm><fnm>A</fnm></au>
    <au><snm>Bartrem</snm><fnm>K</fnm></au>
  </aug>
  <source>Proceedings - 18th IEEE International Conference on Machine Learning
  and Applications, ICMLA 2019</source>
  <publisher>IEEE</publisher>
  <pubdate>2019</pubdate>
  <fpage>1348</fpage>
  <lpage>-1352</lpage>
</bibl>

<bibl id="B41">
  <title><p>{Strategically-motivated advanced persistent threat: Definition,
  process, tactics and a disinformation model of counterattack}</p></title>
  <aug>
    <au><snm>Ahmad</snm><fnm>A</fnm></au>
    <au><snm>Webb</snm><fnm>J</fnm></au>
    <au><snm>Desouza</snm><fnm>KC</fnm></au>
    <au><snm>Boorman</snm><fnm>J</fnm></au>
  </aug>
  <source>Computers {\&} Security</source>
  <publisher>Elsevier Ltd</publisher>
  <pubdate>2019</pubdate>
  <volume>86</volume>
  <fpage>402</fpage>
  <lpage>-418</lpage>
  <url>https://doi.org/10.1016/j.cose.2019.07.001
  https://linkinghub.elsevier.com/retrieve/pii/S0167404818310988</url>
</bibl>

<bibl id="B42">
  <title><p>{Insider Threat Detection Using Log Analysis and Event
  Correlation}</p></title>
  <aug>
    <au><snm>Ambre</snm><fnm>A</fnm></au>
    <au><snm>Shekokar</snm><fnm>N</fnm></au>
  </aug>
  <source>Procedia Computer Science</source>
  <publisher>Elsevier Masson SAS</publisher>
  <pubdate>2015</pubdate>
  <volume>45</volume>
  <issue>C</issue>
  <fpage>436</fpage>
  <lpage>-445</lpage>
  <url>http://dx.doi.org/10.1016/j.procs.2015.03.175
  https://linkinghub.elsevier.com/retrieve/pii/S1877050915004184</url>
</bibl>

<bibl id="B43">
  <title><p>Big Data Analytics for Security</p></title>
  <aug>
    <au><snm>Cardenas</snm><fnm>Aa</fnm></au>
    <au><snm>Manadhata</snm><fnm>PK</fnm></au>
    <au><snm>Rajan</snm><fnm>SP</fnm></au>
  </aug>
  <source>IEEE Security {\&} Privacy</source>
  <pubdate>2013</pubdate>
  <volume>11</volume>
  <issue>6</issue>
  <fpage>74</fpage>
  <lpage>-76</lpage>
  <url>http://ieeexplore.ieee.org/lpdocs/epic03/wrapper.htm?arnumber=6682971
  http://ieeexplore.ieee.org/document/6682971/</url>
</bibl>

</refgrp>
} 

    \end{backmatter}
\end{document}